\documentclass[twoside,twocolumn,9pt]{article}
\usepackage{extsizes}
\usepackage[super,sort&compress,comma]{natbib} 
\usepackage[version=3]{mhchem}
\usepackage[left=1.5cm, right=1.5cm, top=1.785cm, bottom=2.0cm]{geometry}
\usepackage{balance}
\usepackage{mathptmx}
\usepackage{sectsty}
\usepackage{graphicx} 
\usepackage{lastpage}
\usepackage[format=plain,justification=justified,singlelinecheck=false,font={stretch=1.125,small,sf},labelfont=bf,labelsep=space]{caption}
\usepackage{float}
\usepackage{fancyhdr}
\usepackage{fnpos}
\usepackage[english]{babel}
\addto{\captionsenglish}{%
  
}
\usepackage{array}
\usepackage{droidsans}
\usepackage{charter}
\usepackage[T1]{fontenc}
\usepackage[usenames,dvipsnames]{xcolor}
\usepackage{setspace}
\usepackage[compact]{titlesec}
\usepackage{hyperref}

\usepackage{comment}
\usepackage{comment}
\newcommand{\CXP}[1]{\textcolor{black}{#1}}

\newcommand{\MDGrevise}[1]{\textcolor{black}{#1}}

\newcommand{\Ca}{\mathit{Ca}}
\newcommand{\Rey}{\mathit{Re}}

\usepackage{epstopdf}%This line makes .eps figures into .pdf - please comment out if not required.

\definecolor{cream}{RGB}{222,217,201}

\begin{document}

\pagestyle{fancy}
\thispagestyle{plain}
\fancypagestyle{plain}{
%%%HEADER%%%
\renewcommand{\headrulewidth}{0pt}
}
%%%END OF HEADER%%%

%%%PAGE SETUP - Please do not change any commands within this section%%%
\makeFNbottom
\makeatletter
\renewcommand\LARGE{\@setfontsize\LARGE{15pt}{17}}
\renewcommand\Large{\@setfontsize\Large{12pt}{14}}
\renewcommand\large{\@setfontsize\large{10pt}{12}}
\renewcommand\footnotesize{\@setfontsize\footnotesize{7pt}{10}}
\makeatother

\renewcommand{\thefootnote}{\fnsymbol{footnote}}
\renewcommand\footnoterule{\vspace*{1pt}% 
\color{cream}\hrule width 3.5in height 0.4pt \color{black}\vspace*{5pt}} 
\setcounter{secnumdepth}{5}

\makeatletter 
\renewcommand\@biblabel[1]{#1}            
\renewcommand\@makefntext[1]% 
{\noindent\makebox[0pt][r]{\@thefnmark\,}#1}
\makeatother 
\renewcommand{\figurename}{\small{Fig.}~}
\sectionfont{\sffamily\Large}
\subsectionfont{\normalsize}
\subsubsectionfont{\bf}
\setstretch{1.125} %In particular, please do not alter this line.
\setlength{\skip\footins}{0.8cm}
\setlength{\footnotesep}{0.25cm}
\setlength{\jot}{10pt}
\titlespacing*{\section}{0pt}{4pt}{4pt}
\titlespacing*{\subsection}{0pt}{15pt}{1pt}
%%%END OF PAGE SETUP%%%

%%%FOOTER%%%
\fancyfoot{}
\fancyfoot[LO,RE]{\vspace{-7.1pt}\includegraphics[height=9pt]{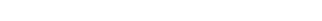}}
\fancyfoot[CO]{\vspace{-7.1pt}\hspace{13.2cm}\includegraphics{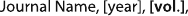}}
\fancyfoot[CE]{\vspace{-7.2pt}\hspace{-14.2cm}\includegraphics{head_foot/RF}}
\fancyfoot[RO]{\footnotesize{\sffamily{1--\pageref{LastPage} ~\textbar  \hspace{2pt}\thepage}}}
\fancyfoot[LE]{\footnotesize{\sffamily{\thepage~\textbar\hspace{3.45cm} 1--\pageref{LastPage}}}}
\fancyhead{}
\renewcommand{\headrulewidth}{0pt} 
\renewcommand{\footrulewidth}{0pt}
\setlength{\arrayrulewidth}{1pt}
\setlength{\columnsep}{6.5mm}
\setlength\bibsep{1pt}
%%%END OF FOOTER%%%

%%%FIGURE SETUP - please do not change any commands within this section%%%
\makeatletter 
\newlength{\figrulesep} 
\setlength{\figrulesep}{0.5\textfloatsep} 

\newcommand{\topfigrule}{\vspace*{-1pt}% 
\noindent{\color{cream}\rule[-\figrulesep]{\columnwidth}{1.5pt}} }

\newcommand{\botfigrule}{\vspace*{-2pt}% 
\noindent{\color{cream}\rule[\figrulesep]{\columnwidth}{1.5pt}} }

\newcommand{\dblfigrule}{\vspace*{-1pt}% 
\noindent{\color{cream}\rule[-\figrulesep]{\textwidth}{1.5pt}} }

\makeatother
%%%END OF FIGURE SETUP%%%

%%%TITLE, AUTHORS AND ABSTRACT%%%
\twocolumn[
  \begin{@twocolumnfalse}
{\includegraphics[height=30pt]{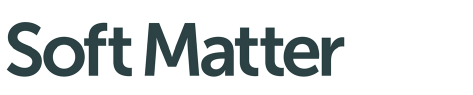}\hfill\raisebox{0pt}[0pt][0pt]{\includegraphics[height=55pt]{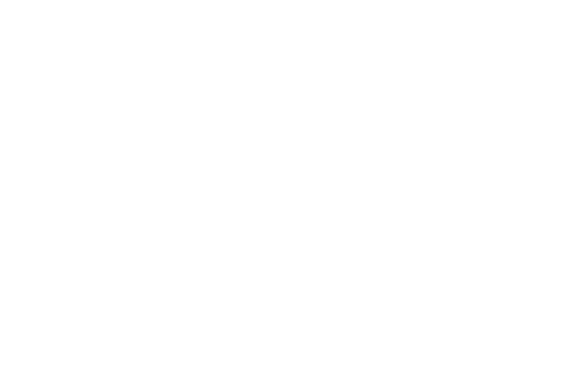}}\\[1ex]
\includegraphics[width=18.5cm]{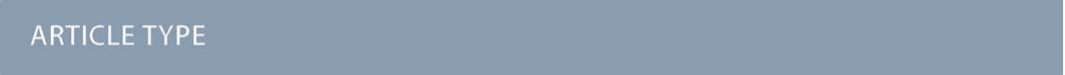}}\par
\vspace{1em}
\sffamily
\begin{tabular}{m{4.5cm} p{13.5cm} }

\includegraphics{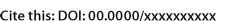} & \noindent\LARGE{\textbf{Red blood cell partitioning and segregation through vascular bifurcations in a model of sickle cell disease}} \\%Article title goes here instead of the text "This is the title"
\vspace{0.3cm} & \vspace{0.3cm} \\

 & \noindent\large{Xiaopo Cheng, \textit{$^{a}$} Christina Caruso\textit, {$^{b}$} Wilbur A. Lam, \textit{$^{bc}$} and Michael D. Graham$^{\ast}$\textit{$^{a}$} } \\%Author names go here instead of "Full name", etc.

\includegraphics{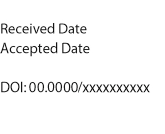} & \noindent\normalsize{The impact of cell segregation and margination in blood disorders on microcirculatory hemodynamics within bifurcated vessels are physiologically significant, yet poorly understood. This study presents a comprehensive computational investigation of red blood cell (RBC) suspension dynamics, with a focus on a model of sickle cell disease (SCD) as an example of a disorder associated with subpopulations of aberrant RBCs. The findings reveal how cell margination influences cellular partitioning and distributions as well as vessel wall shear stress (WSS) at vascular bifurcations. Normal RBCs, which migrate toward the channel center, exhibit the Zweifach-Fung effect, preferentially entering high-flow-rate branches. In contrast, sickle cells, which marginate near the vessel wall, demonstrate an \textit{anti}-Zweifach-Fung effect, favoring lower-flow-rate branches due to their position within the cell-free layer (CFL). The upstream segregation of cells remains downstream through the bifurcation, where sickle cells accumulate along the outer branch walls. This accumulation of sickle cells increases the frequency of high WSS events via direct physical interactions, particularly on the outer side of high-velocity branches, potentially contributing to the vascular damage and endothelial disruption observed in many disorders that affect RBCs. In geometrically asymmetric bifurcations, cells preferentially enter branches with larger radii, underscoring the influence of geometric complexity on microcirculatory blood flow. These findings provide insights into microvascular hemodynamics in SCD  and other blood disorders.} \\%The abstrast goes here instead of the text "The abstract should be..."

\end{tabular}

 \end{@twocolumnfalse} \vspace{0.6cm}

  ]
%%%END OF TITLE, AUTHORS AND ABSTRACT%%%

%%%FONT SETUP - please do not change any commands within this section
\renewcommand*\rmdefault{bch}\normalfont\upshape
\rmfamily
\section*{}
\vspace{-1cm}

%%%FOOTNOTES%%%

\footnotetext{\textit{$^{a}$~Department of Chemical and Biological Engineering, University of Wisconsin-Madison, Madison, WI 53706, USA; E-mail: mdgraham@wisc.edu}}
\footnotetext{\textit{$^{b}$~Aflac Cancer and Blood Disorders Center of Children's Healthcare of Atlanta, Department of Pediatrics, Emory University School of Medicine, Atlanta, GA 30307, USA}}
\footnotetext{\textit{$^{c}$~Wallace H. Coulter Department of Biomedical Engineering, Georgia Institute of Technology and Emory University, Atlanta, GA 30332, USA}}

%%%END OF FOOTNOTES%%%

%%%MAIN TEXT%%%%
%The main text of the article\cite{Mena2000} should appear here.

The partitioning of blood plasma and cells near a vascular bifurcation is complex. In the microcirculation, when red blood cells (RBCs) flow through a bifurcating region of a blood vessel, they tend to preferentially flow into the daughter vessel with the higher flow rate. This phenomenon, called the Zweifach-Fung effect \cite{fung1971microcirculation, vawter1974distribution}, assumes a pivotal role in shaping the distribution of blood flow within the microvascular network. In blood disorders, some RBCs exhibit alterations in their physical properties due to pathological factors, rendering them stiffer and smaller than healthy RBCs, consequently promoting their margination towards blood vessel walls. Blood disorders frequently result in damage to the endothelial cells lining the blood vessel walls, potentially resulting in conditions such as stroke or atherosclerosis\CXP{\cite{caruso2019stiff, caruso2024less}}. Thus, there is a critical need to understand the flow behavior of RBCs in blood disorders and to understand the interplay between cell margination and complex geometries such as bifurcations. \CXP{In this study, we employ detailed computational modeling to investigate microcirculatory dynamics and statistical distributions of blood cells near vascular bifurcations, with a focus on sickle cell disease (SCD). The analysis elucidates how cell segregation and margination influence the flow dynamics of cells through bifurcations,  shaping the wall shear stress (WSS) environment within these vascular structures.}

The spatial distribution of blood components is intricate, influenced by their physical properties such as shape, size, and deformability. In the microcirculation, normal RBCs migrate toward the vascular center, creating a cell-free layer (CFL) depleted of RBCs near the vessel walls. In contrast, white blood cells and platelets preferentially concentrate within the CFL, a flow-induced segregation phenomenon known as margination. Simulations by Kumar et al.\ \cite{Kumar:2011dd, Kumar:2013tu} revealed that stiffer capsules exhibit substantial margination, while softer capsules concentrate near the channel centerline. Similarly, Sinha et al.\ \cite{sinha2016shape} investigated segregation in binary suspensions of spherical capsules, observing that variations in aspect ratio significantly influence their spatial distribution. These findings have been further corroborated through direct simulations of model blood cell suspensions\cite{Qi:2017bf, Mueller:2016fl}. Consistent with the simulation results,  a theoretical framework, based on cell-cell collisions and hydrodynamic migration \cite{Kumar:2012ie, Narsimhan:2013jk, Kumar:2013tu, rivera2015margination,HenriquezRivera:2016wb}, predicts the overall features of RBC suspensions, including the dependence of the CFL thickness on hematocrit and blood vessel size, and also indicates that in suspensions primarily composed of deformable RBCs, rigid cells will strongly concentrate near the walls during flow.

\CXP{Aberrant RBCs arising from blood disorders often exhibit altered physical properties compared to healthy RBCs. A typical example is SCD, where abnormal sickle hemoglobin polymerizes within RBCs under deoxygenated conditions, forming long fibers that disrupt cellular architecture \cite{bertles1968irreversibly}. This pathological process increases membrane stiffness and induces cellular dehydration, leading to reduced cell volume. Consequently, sickle RBCs become less deformable than normal cells, with certain subpopulations irreversibly adopting a characteristic sickle-like shape. In SCD RBC populations, Claver\'ia et al.\ \cite{claveria2021vitro} observed a heterogeneous distribution, with low-density (less stiff) sickle cells preferentially remaining closer to the center of microfluidic channels, while the stiffest cells segregated toward the channel walls.}

\CXP{To investigate the role of cellular interactions in hematological diseases, Caruso et al.\ \cite{caruso2019stiff, caruso2024less} developed an \textit{in vitro} microvasculature model composed of endothelial cells cultured on the inner surface of a microfluidic system. SCD RBCs were introduced into a suspension of normal RBCs and perfused through this endothelialized microfluidic system. The study revealed that the sickled cells marginate, and that VCAM-1, a biomarker of endothelial dysfunction, was upregulated in response to the flow of SCD RBCs compared to normal RBCs. These findings suggest that purely physical interactions between endothelial cells and SCD RBCs, facilitated by margination, are sufficient to trigger endothelial inflammation. Additionally, theoretical and computational studies have demonstrated that diseased cells marginate toward blood vessel walls due to their smaller size and increased stiffness, increasing the frequency of high WSS events \cite{Zhang:2020jt,Caruso.2022.10.1016/j.isci.2022.104606, caruso2024less, Cheng.2023.10.1126/sciadv.adj6423}. These observations underscore the potential for marginated aberrant cells to contribute greatly to the vasculopathy observed in various hematological disorders.}  \MDGrevise{Similar experimental and computational observations are found with iron deficiency anemia (IDA) \cite{Caruso.2022.10.1016/j.isci.2022.104606}, and a detailed computational study suggests that enhanced probability of high WSS events may be found broadly in disorders that generate a population of aberrant RBCs \cite{Cheng.2023.10.1126/sciadv.adj6423}.}

\CXP{The partitioning of cells in blood flow through vascular bifurcations is primarily attributed to the nonuniform cross-sectional distribution of cells in the inlet vessel \cite{pries1996biophysical, secomb2017blood}. Doyeux et al.\ \cite{doyeux2011spheres} quantitatively elucidated the Zweifach-Fung effect, demonstrating that particle enrichment in the high-flow-rate branch arises largely from the initial inlet distribution, which is determined by particle deformability and wall depletion effects \cite{HenriquezRivera:2015fx}. Manoorkar et al.\ \cite{manoorkar2018suspension} investigated the flow of rigid particle suspensions through bifurcating channels, highlighting complex behaviors driven by uneven distributions caused by particle migration in the inlet branch. Using particle tracking velocimetry, Li et al.\ \cite{li2022experimental} analyzed particle transport and downstream concentrations in a successively branching vessel network. Similarly, \textit{in vitro} experiments by Clavica et al.\ \cite{clavica2016red} confirmed the tendency of RBCs to preferentially enter the daughter branch with a higher flow rate. Building on this phenomenon, microfluidic devices have been developed to effectively separate and enrich blood components. For example, Yang et al.\ \cite{yang2006microfluidic} designed a microfluidic device that efficiently separates plasma from blood cells by leveraging the Zweifach-Fung effect.}

\CXP{The hemodynamics of blood cells near vascular bifurcations have been extensively studied. Both \textit{in vivo} experiments and computational simulations \cite{rashidi2023red, kihm2021lingering, pskowski2021investigation, lykov2015inflow, yin2013multiple, leble2011asymmetry10.1063/1.3672689} have documented lingering and jamming effects on RBC partitioning, highlighting the influence of cellular membrane deformability, vascular geometry, and flow conditions. Direct numerical simulations have been instrumental in uncovering the dynamics of RBC suspensions in vascular networks. For example, Balogh et al.\ \cite{balogh2018analysis10.1063/1.5024783} identified cellular-scale mechanisms that govern partitioning dynamics in a model microvasculature, while Zhou et al. \cite{zhou2021emergent} revealed downstream RBC concentration heterogeneity upon altering the network's outflow rates. WSS around bifurcations has also been a focus of investigation. Ye et al.\ \cite{ye2016recovery} examined the effects of flow rate and bifurcation disturbances on the development of the CFL and WSS in branching vessels. Additionally, Balogh et al.\ \cite{balogh2019three} analyzed microvascular networks, revealing significant heterogeneity in WSS and WSS gradients, with the highest WSS occurring in precapillary bifurcations and capillaries.}

\CXP{Altered cellular physical properties, such as deformability and aggregation, significantly influence microvascular hemodynamics near bifurcations. Using a computational model of 3D blood flow in a capillary network, Ebrahimi et al.\ \cite{ebrahimi2022computational} demonstrated that increased RBC stiffness alters RBC trafficking dynamics and elevates local WSS near vascular bifurcations. Rashidi et al.\ \cite{rashidi2023cell} experimentally examined the behavior of healthy and artificially stiffened RBCs in a microfluidic T-junction, revealing distinct focusing profiles in the feeding channel for rigid and healthy RBCs that dramatically impact the cell distribution in the successive daughter channels. In a related computational study, Ye et al.\ \cite{ye2019motion10.1063/1.5079836, ye2019red} investigated the aggregation of multiple RBCs at microvessel bifurcations. Their findings showed that RBCs preferentially flow into the branch with higher RBC flux as cells are compactly arranged into rouleaux, which are difficult to break up at high hematocrit levels.}

\CXP{The margination of cellular components near vascular bifurcations has attracted more attention. In \textit{in vitro} experiments, Sugihara-Seki et al.\ \cite{sugihara2021development} investigated margination in microvessel blood flow using a Y-shaped microchannel, and explored the concentration profile of the marginated particles in the daughter channels. 
Similarly, Bacher et al.\ \cite{bacher2018antimargination} explored microparticle and platelet behavior at vessel confluences and bifurcations, identifying a demargination effect following confluences but not at bifurcations, which influenced platelet concentrations. Sun et al.\ \cite{sun2008lattice} examined RBC-WBC interactions in 2D vascular networks digitized from normal and tumor tissues. Their findings revealed that aggregated RBCs interact more strongly with rolling WBCs, transmitting greater forces to the vessel wall through the WBC and amplifying stress fluctuations at bifurcation sites.  
These insights into WBC margination at bifurcations have inspired the design of microfluidic devices featuring bifurcations to efficiently separate cellular components \cite{shevkoplyas2005biomimetic, wei2012microfluidics10.1063/1.4710992}}.

\CXP{The study of cellular dynamics near vascular bifurcations is critically important for advancing knowledge of blood disorders. Enjalbert et al.\ \cite{enjalbert2021compressed} investigated the effects of tumor vessel compression on tissue oxygenation, uncovering abnormal RBC partitioning at bifurcations that contributes to tissue oxygen heterogeneity. In the context of tumor metastasis, the margination and adhesion of tumor cells play a critical role in determining the sites of extravasation. Wang et al.\ \cite{wang2021margination} simulated tumor cell behavior in a microvascular network, finding that tumor cells are more likely to extravasate at bifurcations in regions with slow blood flow and moderate hematocrit levels. Similarly, Li et al.\ \cite{li2012blood} computationally studied the motion of RBC suspensions in a Y-shaped microfluidic channel, comparing healthy and malaria-infected RBCs. Their findings revealed that the increased stiffness of malaria-infected RBCs causes them to preferentially enter the low-flow-rate daughter branch, a behavior attributed to their distinct initial distribution in the parent channel.}

While numerous experiments and simulations have been conducted to investigate cell flow dynamics in the vicinity of bifurcations, there is a scarcity of studies specifically addressing the impact of margination on cell distribution,  segregation, and wall shear stress at vascular bifurcations. In this study, we employ detailed computational modeling to investigate the dynamics of RBC suspensions in a simple model of sickle cell disease during flow through vascular bifurcations. Our focus encompasses the influence of cell margination on the distribution and partitioning of various cellular components, thereby affecting the stress exerted on the vessel surface. The phenomenon of margination gives rise to a distinct spatial distribution of cells across blood vessels, consequently leading to variations in the partitioning behavior and statistical distribution of different cellular components near bifurcations, and thus to alterations of the Zweifach-Fung effect in the microcirculation.

\section*{Formulation}

We simulate a suspension of RBCs, modeled as deformable fluid-filled elastic capsules, flowing through bifurcated vascular geometries. \CXP{We employ the computational model developed in our previous work; further details are available in \cite{Cheng.2023.10.1126/sciadv.adj6423}}. The vascular system under consideration exhibits a radius ranging from $10$ to $16 \mu m$, with the bifurcating branches geometries conforming to physiological characteristics \cite{murray1926physiological, zamir1978nonsymmetrical}.
The RBC suspensions in our simulations are binary mixtures containing both normal RBCs and aberrant RBCs associated with blood disorders such as sickle cells in SCD. The binary composition assumes a number fraction of $0.9$ for normal RBCs and $0.1$ for aberrant RBCs. This is a substantial simplification, given that real blood cell populations in SCD exhibit a diverse distribution of cellular physical properties. A suspension of only normal RBCs, denoted as a healthy RBC suspension, is considered a control. The overall volume fraction near the inlet (tube hematocrit) is around $20 \%$, consistent with the physiological value in the microcirculation \CXP{\cite{Klitzman.1979.10.1152/ajpheart.1979.237.4.h481, Sarelius.1982.10.1152/ajpheart.1982.243.6.h1018}}.

The distributions of cells and the cell-free layers near the inlet vessels are fully developed before entering the bifurcation region. \CXP{To achieve this, our model incorporates the inflow and outflow boundary conditions for particulate suspensions, as detailed in \cite{liu2020heterogeneous}. 
This boundary condition is particularly well-suited for addressing complex vascular geometries that do not exhibit simple periodic boundary conditions. The implementation involves defining single-phase flow zones at both the inlet and outlet regions. Dirichlet velocity boundary conditions are enforced at the inlet and outlet to establish a Poiseuille flow profile and regulate the relative flow rates in each daughter branch. Downstream of the inlet single-phase zone, a particle introduction zone is established, while upstream of the outlet single-phase zone, a particle removal zone is introduced. Within the particle introduction zone, cells undergo periodic flow. Upon exiting the introduction zone, cells transition into the primary flow domain, at which point an identical cell is introduced at the entrance of the particle introduction zone. Similarly, cells nearing the outlet of the removal zone are continuously removed. This approach ensures that the total number of cells in the simulation is dynamic, eventually converging to a steady range. In addition, the sizes of the single-phase zones and the particle introduction/removal zones are designed to be large enough to facilitate fully developed flow conditions without affecting the flow condition near bifurcations.}

The suspending fluid, blood plasma, is considered incompressible and Newtonian with a viscosity of about $\eta=1.10-1.35 \mathrm{mPas}$. \CXP{In this study, we assume a viscosity ratio of 1 between the intercellular matrix and plasma. While normal RBCs typically exhibit viscosity ratios of up to 15 \cite{recktenwald2022red}, aberrant RBCs associated with various blood disorders can display even higher values. However, previous studies have demonstrated that the dynamics of individual cells \cite{Zhang:2019cl} and cell suspensions \cite{Reasor:2012ey} remain qualitatively robust across a broad range of viscosity ratios.} The discoid radius $a$ for human $\mathrm{RBC}$ is about $4 \mu \mathrm{m}$. The $\mathrm{RBC}$ membrane in-plane shear elasticity modulus $G \sim 2.5-6 \mu \mathrm{N} / \mathrm{m}$. \CXP{In this study, a normal RBC is modeled as a flexible capsule with a spontaneous biconcave discoidal shape representing shear elasticity and an oblate spheroidal shape for bending elasticity \cite{Sinha:2015wt, EVANS:1972uf}. In the context of SCD, abnormal sickle hemoglobin polymerizes within RBCs upon deoxygenation, forming elongated fibers that disrupt the cellular architecture \cite{bertles1968irreversibly}. This pathological process increases membrane stiffness and induces cellular dehydration, thereby reducing the overall cell volume. Consequently, sickle RBCs exhibit significantly reduced deformability compared to normal cells, with certain subpopulations irreversibly adopting a sickle-like shape. In this study, a sickle cell is modeled as a stiff capsule with a curved prolate spheroidal rest shape, with a volume approximately $20\%$ that of the normal RBC model.}The deformability of a capsule in the pressure-driven flow is measured by the dimensionless capillary number $\Ca=\eta \dot{\gamma}_w a / G$. This is set to be $1.0$ for normal RBCs, which corresponds to $\dot{\gamma}_w \sim 1000 \mathrm{~s}^{-1}$. \CXP{The wall shear stress is dimensionless using $\dot{\gamma}_w \sim 1000 \mathrm{~s}^{-1}$.} Since aberrant RBCs in blood disorders are generally much smaller and stiffer than normal RBCs, the interfacial shear modulus $G$ of the aberrant RBCs is assumed to be five times that of normal RBCs, which leads to $C a$ for aberrant RBCs being at most 0.2 times that of $\mathrm{Ca}$ for normal RBCs. All results are for simulations that have been run to a statistically stationary state.

In our simulation, we maintain a particle Reynolds number, denoted as $\Rey_p=\rho \dot{\gamma}{w} a^{2} / \eta$, at a fixed value of $0.1$. \CXP{Although this value of Reynolds number $Re_p$ may be considered relatively high for microcirculation, it represents a compromise between computational accuracy and efficiency. To ensure the robustness of our findings, we performed verification using $Re = 0.05$ and observed no changes in our conclusions.} The fluid is assumed to be incompressible and Newtonian, satisfying the Navier-Stokes equations. Employing a projection method facilitates the temporal advancement of the velocity field $\boldsymbol{u}$. \CXP{The Chorin projection method is utilized to advance the velocity field $\boldsymbol{u}$. This method involves solving an advection-diffusion equation to determine the intermediate velocity field $\boldsymbol{u}^*$, followed by solving a Poisson equation for pressure $P$ to enforce the divergence-free}. The bifurcated vessel is situated within a cuboidal computational domain of $30 a \times 10 a \times 28 a$, and an Eulerian grid with dimensions of $300 \times 100 \times 280$ is employed. Fluid-structure interaction is addressed using the immersed boundary method (IBM). Specifically, our model incorporates two types of immersed boundaries: deformable moving cellular membranes and rigid nonmoving vascular walls. The capsule membrane is discretized into $N_{\Delta}$ piecewise flat triangular elements; $N_{\Delta p}=1280$ for normal RBC, while $N_{\Delta t}=682$ for sickle RBC. Differing values of $N_{\Delta}$ are selected to ensure comparable sizes of triangular elements \CXP{with characteristic size $\sim 0.4 \mu m$} on both capsules. We employ the ``continuous forcing'' 
IBM and ``direct forcing'' IBM methods for the RBC membranes and tube wall, respectively. The numerical methodology follows the approach delineated in previous works \cite{Cheng.2023.10.1126/sciadv.adj6423, balogh2017computational, mittal2008versatile}.

\section*{Results}
\subsection*{Validation: Zweifach-Fung effect}

\begin{figure}[h]
    \centering
    \includegraphics[width=\linewidth]{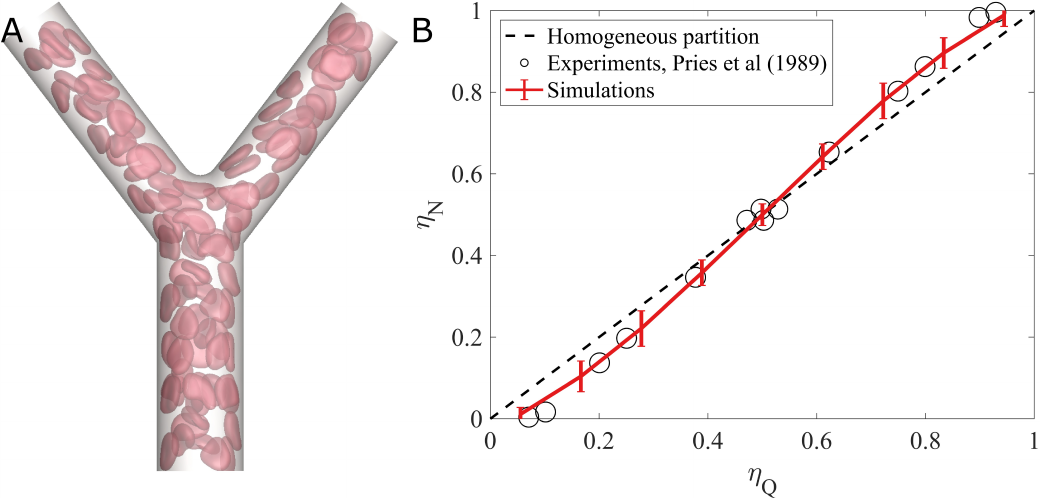}
    \caption{ (A) Simulation snapshots of healthy RBCs partition flowing via a \CXP{geometrically symmetric} bifurcation. The radius of the inlet vessel is set to be $10 \mu m$\CXP{, while the outlet vessel radius is $8 \mu m$.}  
    Cells flow from the bottom inlet to the top outlet branch. (B) Comparison of the Zweifach-Fung effect in our simulations with experimental results  \cite{pries1989red}.}
    \label{fig:val1}
\end{figure}

We begin the description of results with a comparison between our model and experimental data on the Zweifach-Fung effect that is available in the literature \cite{pries1989red}. \CXP{Detailed validation results have been reported elsewhere \cite{Cheng.2023.10.1126/sciadv.adj6423}.} To closely align with physiological parameters employed in experiments, we set the radius of the inlet vessel to $10 \mu m$, as depicted in Fig.~\ref{fig:val1}(A).  To quantify fluid partition within the bifurcation, we introduce the parameter $\eta_Q = Q_{\text{left}} / (Q_{\text{left}} + Q_{\text{right}})$, representing the ratio of the volumetric flow rate at one daughter branch to that at the parent vessel, where $Q_{\text{left}}$ and $Q_{\text{right}}$ denote the volumetric flow rates in the left and right daughter branches, respectively. A higher value of $\eta_Q$ signifies a preference for a greater fluid flow through the left branch. Similarly, the partition ratio for cell number flow rate, denoted as $\eta_N = N_{\text{left}} / (N_{\text{left}} + N_{\text{right}})$, is defined. As $\eta_N$ increases, there is a discernible bias in cell flow toward the left branch as opposed to the right branch within the bifurcation.

As depicted in Fig.~\ref{fig:val1}(B), our simulation results exhibit good agreement with findings from the literature \cite{pries1989red}. Specifically, a consistent pattern emerges wherein, when the left branch experiences a larger outlet volumetric flow rate than the right branch ($\eta_Q > 0.5$), the cell number partition ratio $\eta_N$ exceeds the fluid partition ratio $\eta_Q$, and conversely when the right branch dominates in outlet flow, $\eta_N$ is lower than $\eta_Q$. 

\subsection*{Partitioning of Healthy and Aberrant Cells at a Geometrically Symmetric Bifurcation}

We now characterize the Zweifach-Fung effect in \CXP{the context of SCD}, by considering a mixture suspension of normal RBCs with sickle RBCs flowing through a symmetric bifurcation. The radius of the parent and daughter vessels are set to be $16 \mu m$ and $12.7 \mu m$ respectively, and the angle between the two daughter branches is around $75^{\circ}$, \CXP{as suggested by Murray's law \cite{murray1926physiological} for bifurcating radii and Zamir's law \cite{zamir1978nonsymmetrical} for bifurcation angles} in the circulation. 

\begin{figure*}[h]
    \centering
    \includegraphics[width=\linewidth]{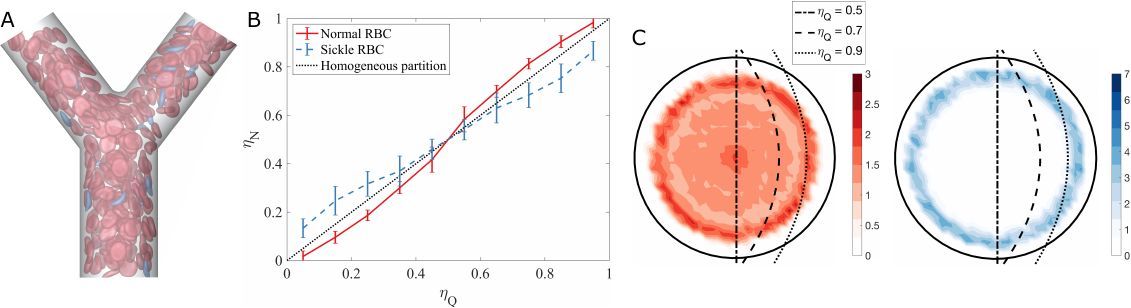}
    \caption{(A) Simulation snapshot of cell partition near a vascular bifurcation in sickle cell disease. Red particles are normal RBCs, while blue are sickle RBCs. Cells flow from bottom to top. \CXP{The bifurcation exhibits geometric symmetry, characterized by a parent vessel radius of $16\,\mu\text{m}$ and daughter vessel radii of $12.7\,\mu\text{m}$. Despite this geometric symmetry, the fluid flow is asymmetric, with a fluid partition ratio of $\eta_Q = 0.9$.} (B) Cell number partition ratio $\eta_N$ as a function of volumetric flow partition ratio $\eta_Q$ for normal and sickle RBCs in a suspension of SCD flowing through the geometrically symmetric bifurcation. (C) Cross-sectional cell number distribution for normal (red) and sickle cells (blue) before the bifurcation and the separatrix at various fluid partition ratios $\eta_Q = 0.5, 0.7, 0.9$. \CXP{The color scale represents the cell number density of normal and sickle cells. In an ideal scenario where the cells are evenly distributed throughout the blood vessels, the cell density would uniformly equal 1 everywhere.}}
    \label{fig:partition}
\end{figure*}

\CXP{As illustrated in Fig.~\ref{fig:partition}(A), the fluid partition ratio, $\eta_Q$, is set to $0.9$. The snapshot depicts the spatial distribution of normal and sickle cells in the vicinity of the bifurcation. Interestingly, a higher concentration of sickle cells is observed in the right branch, which corresponds to the lower volumetric flow rate—a behavior that deviates from the classical Zweifach-Fung effect.} 
\CXP{Some early studies also documented deviations from the classical Zweifach-Fung effect under specific conditions. For example, Shen et al. \cite{shen2016inversion} observed an inversion of the Zweifach–Fung effect in suspensions of less deformable RBCs at very low hematocrit, where the hematocrit in the lower flow-rate daughter branch exceeded that of the parent vessel.}

To illustrate this more quantitatively, the variation of $\eta_N$ relative to $\eta_Q$ is depicted in Fig.~\ref{fig:partition}(B) for normal and sickle cells, respectively. For normal cells in the mixture suspension, the Zweifach-Fung effect is again observed, in which normal RBCs tend to enter the higher flow rate branch preferentially. However, sickle cells do not conform to a similar partitioning behavior as observed in the Zweifach-Fung effect. Notably, at $\eta_Q > 0.5$,  $\eta_N$ for sickle cells is observed to be smaller than $\eta_Q$, and conversely at $\eta_Q < 0.5$. This indicates an inverse tendency in cell partition for sickle RBCs in comparison to normal RBCs; specifically, sickle RBCs demonstrate a preference for entering the branch with the lower flow rate as opposed to the higher flow rate branch; this is an \emph{anti}-Zweifach-Fung effect.

\MDGrevise{To illustrate how the healthy and aberrant cells distribute at the bifurcation, Fig.~\ref{fig:partition}(C) shows histograms of local hematocrit at a cross-sectional position well upstream of the bifurcation along with ``separatrices" indicating which cells partition to the left and right branches for $\eta_Q = 0.5, 0.7, 0.9$. \CXP{The separatrix is defined as the boundary line that determines the eventual flow trajectory of cells: cells located to the left of the separatrix are highly likely to flow into the left branch, while those on the right side are directed into the right branch. Separatrices are calculated and identified using polynomial logistic regression with a precision of around 0.97. Since sickle cells are primarily distributed near the blood vessel wall and insufficient data is available for the center of the channel, the cross-sectional position data of both normal and sickle cells is used to calculate the separatrices. Testing revealed that the separatrices accurately describe the partitioning of both normal and sickle cells, indicating that cell partitioning near the bifurcation is predominantly determined by their position.} 
In this figure we clearly see that the healthy cells exhibit a cell-free layer and a hematocrit profile that exhibits a ring of high hematocrit at the edge of the CFL and then another maximum at the centerline; this is consistent with past observations of cell or deformable capsule distributions \cite{Kumar:2011dd, Kumar:2012ie, HenriquezRivera:2016wb, Cheng.2023.10.1126/sciadv.adj6423}. By contrast, the aberrant cells are strongly marginated, residing in the cell-free layer. } 
Under conditions of even partition ($\eta_Q = 0.5$), the separatrix is a straight line through the channel center,  dividing the cross-sectional area into two equal segments. 
\CXP{As $\eta_Q$ increases, the separatrix gradually shifts and bows towards the right side. This shift plays a crucial role in how the position of the separatrix interacts with the uneven distribution of RBCs, giving rise to the Zweifach-Fung effect. In the conventional Zweifach-Fung effect, the shifts and bending of the separatrix toward the right results in the right region encompassing a much larger fraction of the CFL compared to the left. Consequently, a greater proportion of the cell-free plasma is directed into the right branch, reducing the normal cell number partition ratio, $\eta_N$, relative to $\eta_Q$. In contrast, the opposite trend is observed for marginated sickle cells primarily located within the CFL. Consequently, these sickle cells are preferentially directed toward the right branch with a lower flow rate.}

\subsection*{Cell Distribution and Segregation Downstream of Bifurcation}

\begin{figure*}[h]
    \centering
    \includegraphics[width=\linewidth]{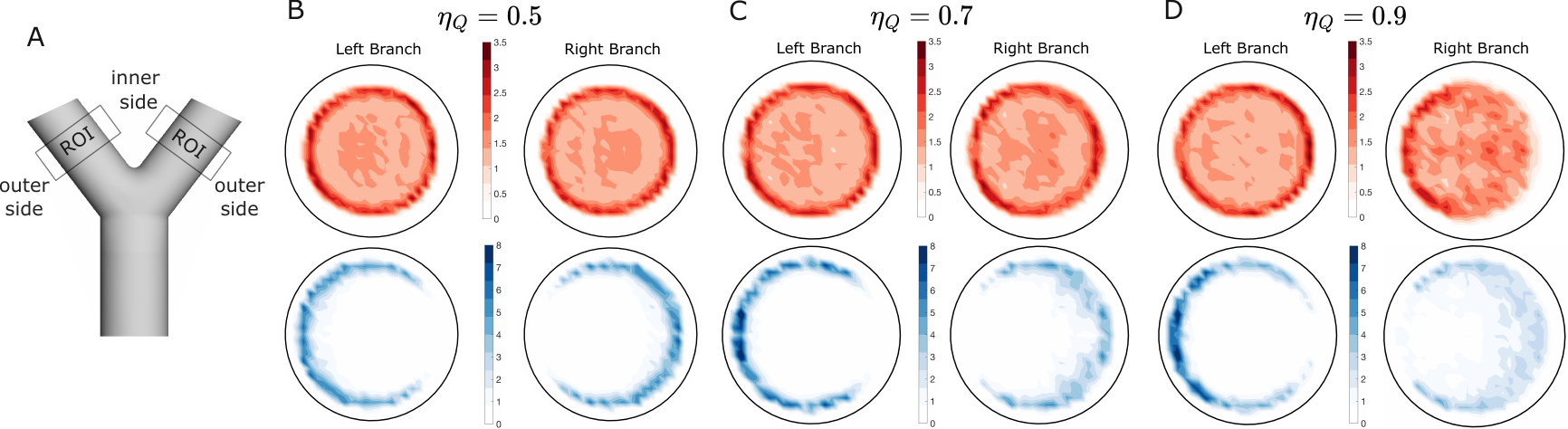}
    \caption{(A) Schematic showing the region of interest (ROI) near the bifurcation; Cross-sectional cell number distribution profile for normal (red) and sickle cells (blue) in the left and right branches downstream the bifurcation at various fluid partition ratios (B) $\eta_Q = 0.5$, (C) $\eta_Q = 0.7$, and (D) $\eta_Q = 0.9$. The color scale represents the cell number density of normal and sickle cells. In an ideal scenario where the cells are evenly distributed throughout the blood vessels, the cell density would uniformly equal 1 everywhere.}
    \label{fig:distribution1}
\end{figure*}

To characterize the distribution of cells after the flow bifurcation, \MDGrevise{Fig.~\ref{fig:distribution1} shows the cross-sectional distributions of number density downstream of the bifurcation, averaged over the ``regions of interest" (ROI) indicated in the figure.} \CXP{This study examines the influence of bifurcation geometry on downstream cell distribution and segregation. Under uniform partitioning conditions $(\eta_Q = 0.5)$, the cross-sectional cell number density in the left and right daughter branches reveals persistent segregation between normal and sickle cells: normal cells predominantly occupy the center of the branches, while sickle cells accumulate near the vessel walls. Notably, downstream of the bifurcation, sickle cells are concentrated along the walls, with a higher density observed on the outer side (away from the bifurcation center) and an absence on the inner side (closer to the bifurcation center). When partitioning becomes biased $(\eta_Q > 0.5)$, increasing $\eta_Q$ leads to higher flow velocities in the left branch and reduced velocities in the right branch. At $\eta_Q = 0.7$, the distribution of normal  RBCs remains largely unaffected, while sickle cells in the left branch exhibit greater concentration near the outer wall, and their distribution in the right branch becomes less concentrated. At $\eta_Q = 0.9$, this trend becomes more pronounced, with normal RBCs in the right branch being biased towards the inner side wall. The cell-free layer on the outside of the right branch also becomes thicker. Simultaneously, the distribution of sickle cells in the right branch becomes less concentrated, due to fewer normal cells on the outside of the right branch.}

\subsection*{Wall Shear Stress on Bifurcated Vessel}

\begin{figure*}[h]
    \centering
    \includegraphics[width=\linewidth]{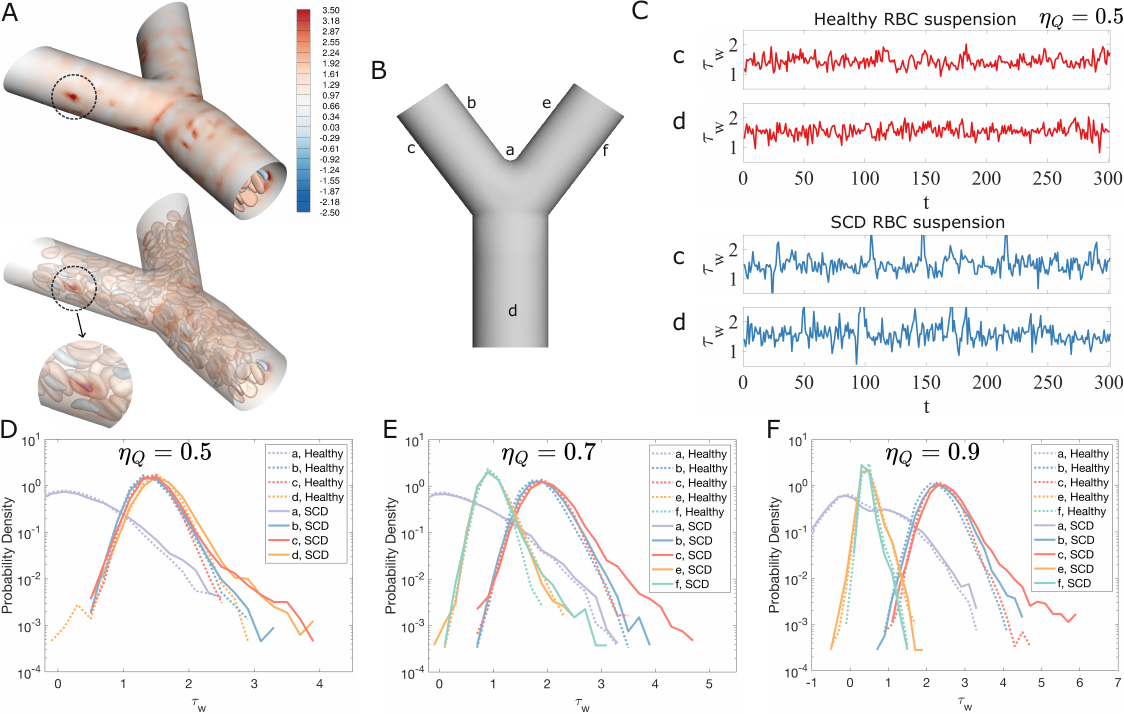}
    \caption{(A) Transient wall shear stress $\tau_w$ on (top) the bifurcation vascular surface and (bottom) corresponding transparent image. An area where a sickle cell close to the wall directly provokes WSS fluctuation in the daughter branch is indicated with a circle. (B) Schematic diagram of several sampling areas $(a-f)$ on the surface of bifurcated blood vessels. (C) Temporal evolution of wall shear stress $\tau_w$ at selected fix points $(c, d)$ on the bifurcation vascular surface with $\eta_Q = 0.5$. 
    (D-F) Probability density profile of WSS $\tau_w$ at selected fix points on the bifurcation vascular surface with (D) $\eta_Q = 0.5$ (E) $\eta_Q = 0.7$, and (F) $\eta_Q = 0.9$.}
    \label{fig:wallshearstress}
\end{figure*}

We have seen that the stiffer aberrant cells exhibit a pronounced tendency to marginate, resulting in close approaches between these RBCs and the blood vessel walls (endothelium). Experimental observations indicate that such interactions induce local fluctuations in wall shear stress, consequently triggering endothelial damage and inflammation \cite{Cheng.2023.10.1126/sciadv.adj6423}. To gain a comprehensive understanding of the hydrodynamic ramifications associated with cell margination and segregation at the vascular surface, we conduct an evaluation of the WSS near the bifurcation. As depicted in Fig.~\ref{fig:wallshearstress} (A), illustrating the transient distribution of WSS on the bifurcated vessel surface, distinct positive local WSS fluctuations are observed on the lateral side of the daughter branch, which is attributed to the proximity of sickle cells to the vessel wall, as evident in the corresponding transparent image.

\CXP{For a detailed characterization of stress evolution over time at different locations on the vessel surface, we strategically selected specific fixed points $(a-f)$ on the blood vessel surface of the bifurcated geometry, as illustrated in Fig.~\ref{fig:wallshearstress}(B). The temporal evolution of WSS at these designated points was analyzed and compared between two suspensions: one containing only healthy RBCs and the other containing SCD RBCs. In Fig.~\ref{fig:wallshearstress}(B), point \(a\) is situated at the center of the bifurcation, while points \(b\), \(c\), \(e\), and \(f\) correspond to the inner and outer sides of the left and right sub-branches, respectively. In addition, point \(d\) represents a location distal to the bifurcation, closer to the inlet. A comparative analysis in Fig.~\ref{fig:wallshearstress}(C) between the healthy and SCD RBC suspensions reveals large differences. The SCD suspension exhibits markedly larger WSS fluctuations, particularly at points \(c\) and \(d\). Frequent and pronounced peaks are observed in the WSS temporal signal for the SCD suspension, whereas these peaks are absent in the healthy suspension. This disparity arises from the propensity of stiff, sickled cells to marginate and approach the vascular wall, inducing substantial local WSS fluctuations, as shown in Fig.~\ref{fig:wallshearstress}(A).}

\CXP{To enhance statistical insight, Fig.~\ref{fig:wallshearstress}(D) illustrates the probability density distribution of WSS under even flow partition (\(\eta_Q = 0.5\)) at selected points on the bifurcation surface for both healthy and SCD RBC suspensions. The distribution at point \(a\) is the broadest, aligning with the largest WSS fluctuations observed at this location. This is primarily due to the persistent presence of cells in proximity to the bifurcation. Despite these fluctuations, the average WSS at point \(a\) remains close to zero, owing to the geometric symmetry of the bifurcation. The presence of sickled cells notably increases the probability of encountering high WSS, particularly at points \(c\) and \(d\), due to the tendency of stiff sickle cells to marginate toward the cell-free layer near the vessel wall and concentrate along the outer sides of daughter branches. This effect is less pronounced at point \(c\), as fewer sickled cells are found on the inner side of the branch. Extending this analysis, Figs.~\ref{fig:wallshearstress}(E, F) present WSS distributions under uneven flow partition (\(\eta_Q > 0.5\)), focusing on five points (\(a\), \(b\), \(c\), \(e\), \(f\)) on the bifurcation surface. The average WSS at points \(b\) and \(c\) exceeded that at \(e\) and \(f\), reflecting the higher flow velocity in the left branch. Notably, the SCD RBC suspension exhibited a higher probability of elevated WSS, particularly at point \(c\), located along the high-velocity branch, attributed to the accumulation of stiff sickled cells near the outer vessel wall. The WSS of the low-velocity branch is less sensitive to the presence of sickle cells, which is perhaps related to the observed reduced segregation. Overall, the margination and flow conditions within the bifurcation geometry influence the downstream distribution of sickle cells, leading to more large wall stress events on the outer side of the branch, where marginated sickle cells are more concentrated.}

\subsection*{Geometrically Asymmetric Bifurcation}

\begin{figure*}[h]
    \centering
    \includegraphics[width=\linewidth]{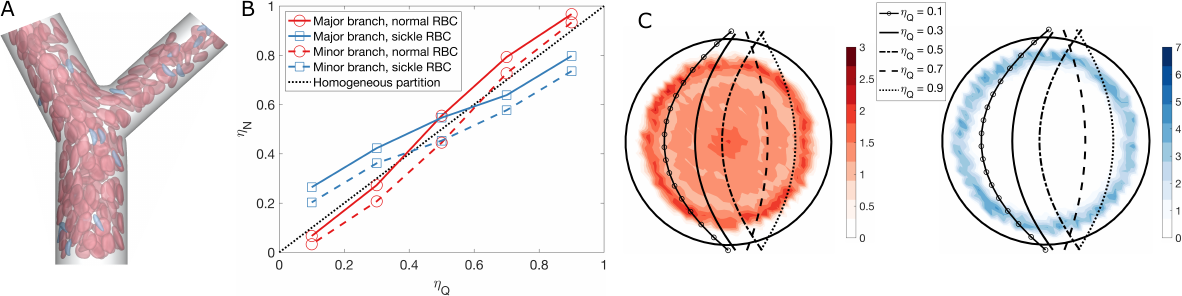}
    \caption{(A) A simulation snapshot of cell partition near a geometrically asymmetric vascular bifurcation in sickle cell diseases with $\eta_{Q, \text{ major}}=0.9$. Cells flow from the bottom inlet to the top outlet branch. In this asymmetric bifurcation geometry, the vessel radius at the entrance is \(16 \, \mu m\), the left major branch has a radius of \(14 \, \mu m\), and the right minor branch has a smaller radius of \(11 \, \mu m\). 
    (B) the cell number partition ratio $\eta_N$ as a function of the volumetric flow partition ratio $\eta_Q$ for normal RBCs (red) and sickle RBCs (blue).  (C) Cross-sectional cell number distribution for normal (red) and sickle cells (blue) before the asymmetric bifurcation and the separatrix at various fluid partition ratios $\eta_{Q, \text{ major}} = 0.1, 0.3, 0.5, 0.7, 0.9$. }
    \label{fig:asymmetry}
\end{figure*}

\CXP{We now present results for an asymmetrical geometry, which is more representative of physiological conditions than the symmetric case considered above. To explore the impact of geometric asymmetry on cellular blood flow, we analyze the behavior of normal and sickle cells within an asymmetric bifurcation. As shown in Fig.~\ref{fig:asymmetry}(A), with \(\eta_{Q}\) set to 0.9 for the left branch (larger radius), a greater accumulation of sickle cells and a thicker cell-free layer are observed in the right branch, which has a smaller radius and a lower volumetric flow rate. \CXP{Note that the definition of partition ratio $\eta$ in this figure is slightly different from the previous ones because of the introduction of geometric asymmetry. $\eta_{\text{major}}$ and $\eta_{\text{minor}}$ are defined for the major and minor branches, respectively. For instance, $\eta_{Q, \text{ minor}} = Q_{\text{minor}}/(Q_{\text{major}} + Q_{\text{minor}})$.} Fig.~\ref{fig:asymmetry}(B) depicts the relationship between \(\eta_N\) and \(\eta_Q\) for both cell types in the left (major) and right (minor) branches, revealing deviations from partitioning patterns observed in symmetric geometries. Even at \(\eta_Q = 0.5\), cells exhibit a preference for the major branch, with \(\eta_N > 0.5\) for the larger-radius branch. This finding emphasizes the critical role of geometric asymmetry in cell distribution, favoring flow into the branch with a larger radius. Moreover, sickle cells again demonstrate the \emph{anti}-Zweifach-Fung effect, exhibiting a tendency to flow preferentially into branches with lower volumetric flow rates.}

\CXP{To further investigate the impact of bifurcation asymmetry on cell partitioning behavior, Fig.~\ref{fig:asymmetry}(C) illustrates the separatrix at the entrance of the bifurcation. Under asymmetric conditions, even when the volumetric flow rates in the major and minor branches are equal (\(\eta_Q = 0.5\)), the separatrix is notably concave rather than straight. This curvature arises due to the larger radius of the left branch, resulting in \(\eta_N > 0.5\) when \(\eta_Q = 0.5\), which highlights a preferential flow of cells into the branch with the larger radius. The influence of geometric asymmetry on sickle cells is similar to that on normal cells; however, due to the proximity of marginated sickle cells to the vessel wall and their tendency to stay within the cell-free layer, sickle cells continue to exhibit the \emph{anti}-Zweifach-Fung effect. Overall, the geometric asymmetry of the bifurcation increases the curvature of the separatrix, further amplifying the uneven distribution of cells within the plasma.}

\section*{Discussion}

\CXP{Our study first addresses the flow of cell suspensions through geometrically symmetric bifurcations, replicating the Zweifach-Fung effect, where cells preferentially flow into high-flow-rate branches. When analyzing a binary mixture modling SCD blood, contrasting behaviors are found: while normal RBCs exhibit the expected Zweifach-Fung effect, stiff sickle cells demonstrate an \textit{anti-Zweifach-Fung} effect, favoring lower-flow-rate branches. This arises because upstream of the bifurcation, normal RBCs are concentrated in the channel center, while marginated sickle cells predominantly reside in the CFL near the vessel wall. Under symmetric geometric and flow partition conditions (\(\eta_Q = 0.5\)), the separatrix dividing upstream regions where healthy RBCs go to the left or right forms a straight line through the channel center but shifts and curves as \(\eta_Q\) increases. This curvature leads to uneven plasma and cell partitioning, with relatively more cell-free plasma entering the branch with a lower volumetric flow rate. Thus, marginated sickle cells, localized mainly within the CFL, exhibit an opposite trend to the Zweifach-Fung effect,  underscoring the distinct partitioning behaviors of normal and aberrant cells in blood disorders.}

\CXP{This study also examines the influence of bifurcation geometry on downstream cell distribution. When a suspension of SCD cells flows through a bifurcated vessel, the segregation between cells observed upstream is preserved and remains downstream. Across a range of \(\eta_Q\) values, normal cells predominantly occupy the central region of the branch, while sickle cells accumulate in the cell-free layer near the vessel wall, particularly along the outer side of the branch. Furthermore, this study investigates the impact of SCD RBC suspensions on WSS in vascular bifurcations. Simulations reveal that marginated stiff sickle cells exhibit increased physical interactions with the vessel wall, leading to pronounced local WSS fluctuations. Statistical analysis of WSS across key regions of the bifurcated vessel surface demonstrates that SCD suspensions greatly increase the probability of high WSS events compared to healthy cell suspensions, particularly near the outer branch wall where sickle cells preferentially accumulate downstream. This effect is further exacerbated in high-velocity branches under uneven flow conditions (\(\eta_Q > 0.5\)). These findings underscore the influence of cell segregation and flow environment in amplifying wall stress fluctuations, which potentially contribute to vascular and endothelial damage associated with sickle cell disease and other types of blood disorders.}

\CXP{Finally, the study explores the impact of geometric asymmetry in vessel bifurcations on cellular blood flow. In asymmetric geometries, where one branch has a larger radius, cells preferentially flow into the larger-radius branch, resulting in $\eta_N>0.5$ even under equal volumetric flow conditions ($\eta_Q=0.5$). Stiff sickle cells still display the {anti-Zweifach-Fung effect}, favoring branches with lower flow rates due to their margination and position in the cell-free layer. Under asymmetric conditions, the separatrix becomes more curved, highlighting the critical role of vascular geometry complexity in shaping blood flow dynamics in the microcirculation.}

\section*{Author Contributions}

M.D.G, W.A.L., and C.C. designed research. X.C. performed research. All authors analyzed data and wrote the paper. 

\section*{Conflicts of interest}
The authors declare that they have no competing interests.

\section*{Data availability}
All data needed to evaluate the conclusions in the paper are present in the paper and/or the Supplementary Materials.

\section*{Acknowledgments}
This work was supported by NSF grant CBET-2042221 and ONR grant N00014-18-1-2865 (Vannevar Bush Faculty Fellowship) (M.D.G., and X.C.), an American Society of Hematology (ASH) Research Training Award for Fellows (RTAF), NIH National Heart, Lung, and Blood Institute (NHLBI) grant T32HL139443 and Pediatric Loan Repayment Program (LRP) Award L40HL149069 (C.C.), and NIH NHLBI grant R35HL145000 (W.A.L.). The work was performed, in part, at the Georgia Tech Institute for Electronics and Nanotechnology, a member of the National Nanotechnology Coordinated Infrastructure (NCCI), which is supported by the NSF grant ECCS-2025462. This work used the Advanced Cyberinfrastructure Coordination Ecosystem: Services \& Support (ACCESS). In particular, it used the Expanse system at the San Diego Supercomputing Center (SDSC) through allocation MCB190100, and PHY240144.

\bibliography{bloodMDG, S24proposalrefs, scibib, ref}
\bibliographystyle{rsc} %the RSC's .bst file

\end{document}